%% file: main.tex
\documentclass[journal, a4paper]{IEEEtran}
\usepackage{cite}

\ifCLASSINFOpdf
  \usepackage[pdftex]{graphicx}
  \graphicspath{{fig/}}
  \DeclareGraphicsExtensions{.pdf,.jpeg,.png}
\else
\fi  

\usepackage{amsmath,amsfonts,amssymb}
\usepackage{algpseudocode}
\usepackage{algorithm}
\usepackage{array}
\usepackage{multirow}

\usepackage{subcaption}	
\usepackage{import}
\usepackage{xcolor}
\usepackage[hidelinks]{hyperref}
\usepackage[squaren]{SIunits}
\usepackage{booktabs}
\usepackage{soul}
\usepackage{wasysym}
\usepackage{bm}
\usepackage{pifont}
\usepackage{mathtools}
\usepackage{tablefootnote}
\usepackage{upgreek}
\usepackage[utf8]{inputenc}
\usepackage[T1]{fontenc}

\usepackage{pgf}
\usepackage{pgfplots}
\pgfplotsset{/pgf/number format/use comma, compat=newest}
\usepgfplotslibrary{external}
\usepackage{multirow}
\usepackage{booktabs}

\usetikzlibrary{arrows.meta}
\usetikzlibrary{calc}
\usetikzlibrary{shapes.geometric, arrows, positioning}

\tikzset{
    block/.style={
        rectangle,
        draw,
        rounded corners,
        minimum height=2cm,
        minimum width=3cm,
        align=center,
        fill=blue!10
    },
    arrow/.style={
        ->,
        thick
    }
}

\tikzset{
  bigblock/.style = {rectangle, draw=black, thick, rounded corners, fill=blue!15, inner sep=0.5cm},
  subblock/.style = {
  rectangle, draw=black, fill=gray!20,
  minimum width=2.7cm, minimum height=0.8cm,
  align=center, text width=2.7cm},
  decision/.style = {diamond, draw=black, fill=gray!20, aspect=2,
	minimum width=4cm, minimum height=1.5cm, align=center, inner sep=1pt},
  arrow/.style = {thick,->,>=stealth}
}

\tikzstyle{startstop} = [ellipse, minimum width=3cm, minimum height=1cm,text centered, draw=black, fill=gray!20]
\tikzstyle{process} = [rectangle, minimum width=6cm, minimum height=1cm, text centered, draw=black, fill=blue!15, text width=6cm, align=center]
\tikzstyle{arrow} = [thick,->,>=stealth]

\tikzset{
  grid step/.initial=6mm,
}

\tikzset{
  pics/grid/.style args={#1/#2}{
    code={
      \pgfkeysgetvalue{/tikz/grid step}{\gridstep}
      \pgfmathsetlengthmacro{\w}{#2*\gridstep}
      \pgfmathsetlengthmacro{\h}{#1*\gridstep}
      \draw[step=\gridstep] (0,0) grid (\w,\h);
    }
  }
}

\tikzset{
  gridbox/.style={draw, minimum width=6mm, minimum height=6mm},
  conv1d/.style={draw, minimum width=6mm, minimum height=6mm},
  statbox/.style={rectangle, draw, minimum width=14mm, minimum height=6mm, align=center},
  startstop/.style={ellipse, draw, minimum width=3cm, minimum height=1cm, align=center, fill=gray!20},
  process/.style={rectangle, draw, align=center, fill=blue!15, text width=4cm, minimum width=4cm, minimum height=1cm},
  decision/.style={diamond, draw, align=center, aspect=2, minimum width=4cm, minimum height=1.5cm, fill=gray!20},
  arrow/.style={thick, ->, >=Stealth},
}

\renewcommand{\figurename}{Fig.}
\renewcommand{\tablename}{Tab.}

\newcommand{\secname}{Sec.}

\newcommand{\figref}[1]{\figurename~\ref{#1}}
\newcommand{\tabref}[1]{\tablename~\ref{#1}}
\newcommand{\secref}[1]{\secname~\ref{#1}}

\newcolumntype{C}[1]{>{\centering\arraybackslash}p{#1}}

\begin{document}
\renewcommand\arraystretch{1.4}
\title{
Container Unloading via Reinforcement Learning: \\
Picking Order, Deadlock Avoidance, and Proof-of-Concept Simulation\\
}

\author{\IEEEauthorblockN{Jan R\"udiger, Max Schenke, Daniel Weber}
\thanks{J.~R\"udiger is studying electrical engineering at Paderborn University, Germany. M.~Schenke and D.~Weber are with the Department of Power Electronics and Electrical Drives at Paderborn University, Germany. E-Mail: \{schenke, weber\}@lea.uni-paderborn.de
}
}
\renewcommand{\abstractname}{Abstract}

\maketitle
\begin{abstract} 
Unloading containers in the courier, express and parcel industry is a physically demanding and labor-intensive work. Automatizing this process is an important step towards increasing the efficiency of parcel-handling systems. This work investigates the potential of reinforcement learning to learn a policy for item selection in container unloading scenarios. For that, a simulation environment is created and a masked deep Q-learning with a specially designed neural network architecture is implemented. The results indicate that the agent can learn to select items with an average success rate of 60\,\%, which is significantly better than a random policy at a random chance of 20\,\%. The findings suggest that RL could be a promising approach for automatizing item unloading tasks in the future. 
\end{abstract}


\IEEEpeerreviewmaketitle

\import{chapters}{Introduction.tex}
\import{chapters}{Fundamentals.tex}
\import{chapters}{Implementation.tex}
\import{chapters}{Results.tex}
\import{chapters}{Summary.tex}
\import{chapters}{Appendix_Derivation.tex}

\bibliographystyle{IEEEtran}
\bibliography{Sources}

\vfill

\end{document}

%% file: chapters/Introduction.tex
\section{Introduction}
\label{sec:introduction}
\IEEEPARstart{I}{n} times of labor shortage and increasing labor costs, automation of handling goods becomes more and more important. 
In the courier, express and parcel (CEP) industry, some labor intensive tasks are still performed manually, such as unloading items from trailers or containers, to feed them into subsequent sorting systems. 
One of its subproblems is to decide which item to pick next. 
This work investigates the potential of reinforcement learning (RL) for the proper selection of unloadable items within a transport container. 

A conventional approach could be to use a heuristic that selects items based on manually defined criteria. 
For that, rules need to be defined, e.g., to always pick an item that is lying on top. 
The problem with rule-based decisions is, that expert knowledge is required to specify formal and sufficient criteria. 
Yet, experience-driven heuristics are not guaranteed to resolve the unloading task in an optimal way, e.g., according to a metric like throughput.
Even though it can be measured how well the decision rule performs, it does not provide an obvious indication how to improve it. 
Further, the potential of heuristic solutions does oftentimes decline with increasing problem complexity.
In idealized scenarios, e.g., properly stacked cardboard boxes that should be picked in a single pick manner, a powerful heuristic can be designed quite easily.
In a more comprehensive scenario, such as selecting the optimal pick from an unordered pile of packages in a bulk pick manner, it becomes difficult to define criteria that lead to good decisions based on a given metric. 
These considerations give rise to the question how optimal decision making could be realized in the context of the described container unloading task.

Abstracting from the physical scenario, the task of picking items can be modeled as a Markov decision process (MDP) and, therefore, the problem can be approached with the tools of RL. 
An RL algorithm could deal with the container unloading environment while directly optimizing its policy based on given spatial and performance measurements. 
This way, it can be targeted to decouple the achievable performance from the available expert knowledge, as decision rules will be inferred and improved upon observations.
Contrary to heuristic solutions, the iterative learning process that is fundamental to RL renders it an (approximate) optimal decision making tool.

In order to assess the general applicability of RL for unloading tasks, this investigation focuses an idealized scenario featuring neatly stacked packages.
Although such a setup would be approachable by means of heuristic methods, it also poses as a well-structured example task for RL as a proof-of-concept solution.
Overall, the simplified container unloading scenario is characterized as follows:
\begin{itemize}
    \item items are stacked in an ordered fashion,
    \item unloading progresses by selecting individual packages to be picked, rendering the setup a single-pick task,
    \item all items have identical mass and friction properties,
    \item spatial information of position, orientation and package size are assumed available, i.e., the upstream retrieval of this information is not considered,
    \item force can be applied directly, i.e., the downstream control of a robot manipulator is not considered,
    \item packages that can be moved by applying force are considered movable, and are then directly deleted from the container.
\end{itemize}

\subsection{Contribution}
This work proposes solving the unloading task using RL by training an agent in a physics-based simulation environment to select optimal items. As this research is at an early stage, a dedicated framework is developed as a fundamental proof of concept. The main contributions can be summarized as follows:
\begin{itemize}
    \item Development of a physics-based simulation environment for the unloading task.
    \item An enhanced deep Q-learning approach incorporating action masking for invalid or infeasible actions.
    \item Design and implementation of suitable state features and feature engineering.
    \item A neural network architecture specifically designed for permutation-invariant feature representations.
    \item Experimental validation demonstrating the feasibility of applying RL to the unloading problem.
\end{itemize}

The article is set up as follows:
a brief overview about the physical simulation will be presented in \secref{sec_Fundamentals}, along with the deep Q-learning (DQN) RL algorithm, which will be used to tackle this task.
In \secref{sec_Implementation}, the specific setup of the training environment is described, before task-specific adaptations to DQN are discussed.

\secref{sec_Results} summarizes the results of this work, and \secref{sec_Conclusion_and_Outlook} concludes the findings delivers a short outlook to potential next steps.

%% file: chapters/Fundamentals.tex
\section{Fundamentals} \label{sec_Fundamentals}

The following discussion provides a brief overview about 3D simulation in the context of RL applications and revisits the DQN algorithm from the domain of deep RL. 

\subsection{Simulation-based Reinforcement Learning}
RL allows the synthesis of an (approximate) optimal policy by improving an initial policy in direct interaction with an environment that contains the problem setting.
Herein, the environment could be either, a physical process or a simulation model that mimics the input and output interface of a physical system.
 
In many applications, simulations are preferred for the training of RL policies, because they usually are significantly more cost-efficient, enable shorter training procedures, and are free of risk with respect to damage to hardware and personnel. 
While some applications are too comprehensive to model, which would prevent a simulation environment from producing feasible policies, the setting that is investigated in this work is physically well-defined and can be simulated with commercially available software.
The simulation requirements can therefore be qualitatively summarized as follows:
\begin{itemize}
    \item the simulator must reproduce the relevant behavior of the physical system with sufficient accuracy, which ensures that the learned policy transfers feasibly to the application task in the real world,
    \item the RL algorithm must be able to interact with the simulation in the same way as with a physical environment, i.e., the application of actions and the retrieval of state measurements must succeed via the same interface in either implementation,
    \item the signals presented within the interface, i.e., the selection of observed features, must provide sufficient information to allow effective decisions in the context of the targeted problem setting.
\end{itemize}

\subsection{Deep Q-Learning}\label{sub_sec_standard_DQN}
The item picking task is modeled as an MDP, defined by the tuple
$\langle \mathcal{X}, \mathcal{U}, \boldsymbol{P}, r, \gamma \rangle$ \cite{SuttonBarto2015},
where $\mathcal{X} \subset \mathbb{R}^n$ denotes the continuous state space,
$\mathcal{U}$ is a finite discrete action space, $\boldsymbol{P}(\boldsymbol{x}_{k+1} \mid \boldsymbol{x}_k, u_k)$ represents the state transition probability, $r(\boldsymbol{x}_k, u_k)$ is the reward function, and $\gamma \in [0,1]$ is the discount factor.\\

This work utilizes the DQN algorithm to provide the action-value $q$ as expected return $g$:
\begin{align}
  q_\pi(\boldsymbol{x}_k, u_k) 
  &= 
  \text{E}_\pi 
  \left[
  g_k | \boldsymbol{x}_k, u_k 
  \right], 
  \label{eq_DQN_Q_val_return}  
\end{align}
with the observation vector $\boldsymbol{x}_k$, the action $u_k$, and expectation operator $\text{E}$ \cite{SuttonBarto2015}. 
The return $g$ is defined as the sum of all upcoming rewards $r$
\begin{align}
    g_k = r_{k+1} + \gamma r_{k+2} + \gamma^2 r_{k+3} + \cdots =   \sum_{i=0}^{\infty}\gamma^i r_{k+i+1}, \label{eq_discounted_return}
\end{align}
with discount factor $\gamma \in [0, 1]$ \cite{SuttonBarto2015}.
Whereas the immediate reward $r$ only assesses the momentary performance with respect to the RL problem, the return $g$ incorporates the future development of the performance and can, hence, be considered for planning tasks.
The discount factor defines the temporal foresight with which future rewards are considered within $g$. 
In the engineering context of RL both, $\gamma$ and $r$, are oftentimes available as design degrees of freedom, which can then be configured to design the task and its foresightedness with respect to the desired outcome.
Assuming knowledge over $q$, a control policy $\pi(\bm{x})$ can be retrieved by selecting and applying the action that maximizes $q$:
\begin{align}
  \pi(\boldsymbol{x}_k) = \underset{u}{\arg\max}\,q_\pi(\bm{x}_k, u),
\end{align}
wherein the notation $q_\pi$ denotes the assumption that the policy $\pi$ is followed consistently \cite{SuttonBarto2015}. 

For the core of RL tasks, it remains to determine $q$ according to \eqref{eq_DQN_Q_val_return}, whereas in the general case, the reward function $r$ is unavailable for analytical investigation and can only be accessed via samples.
Due to the continuous nature of the state vector $\bm{x}$, tabular representation of $q(\bm{x}, u)$ cannot be accomplished without accepting a spatial discretization error.
To avoid corresponding loss of resolution, the DQN algorithm has been proposed \cite{Mnih2015}, which is equipped for dealing with continuous state and finite action spaces by utilizing function approximation.
Herein, the action value function is approximated by means of a parameterized function approximator $\hat{q}(\bm{x}, u, \bm{\theta})$, whereas $\bm{\theta}$ denotes the entirety of necessary parameters.
As approximation method, deep neural networks are a pragmatic choice when nonlinear relations must be assumed, with stateless feedforward architectures being the standard \cite{Mnih2015}. 
That is, estimating the action value function $\hat{q}$ succeeds by adapting the approximator parameters $\bm{\theta}$ on the basis of the data that can be collected in direct interaction with the task environment.
The employed loss function $L(\bm{\theta})$ is the mean squared error (MSE) between the target and the momentary action value estimate:
\begin{align}
  L(\bm{\theta}) 
  = 
  \Big( 
  \underbrace{
  r_{k+1} 
  + 
  \gamma \max_{u} 
  \hat{q}(\boldsymbol{x}_{k+1}, u, \bm{\theta}^-) 
  }_{\text{target}}
  - 
  \hat{q}(\boldsymbol{x}_k, u, \bm{\theta}) 
  \Big)^2, 
  \label{eq_Standard_DQN_loss_fcn_MSE}
\end{align}
where the target is being computed using delayed weights $\hat{q}(\boldsymbol{x}, u, \bm{\theta}^-)$, whereas the momentary action value is calculated using the momentary weights $\hat{q}(\boldsymbol{x}, u, \bm{\theta})$~\cite{Mnih2015}. 
In \secref{sub_sub_sec_masked_DQN}, a detailed description of the neural network used to approximate the action value $q$ and further adaptations are described. 

In order to find an optimal policy, it is further required that the data foundation upon which is being trained contains sufficiently variant information about states and actions. 
Intuitively, the training procedure cannot be expected to yield a specific policy, if it was not possible to try and evaluate this policy during the training phase. 
Hence, it is beneficial to the performance outcome to enforce exploration of different state-action combinations by following a partly randomized policy during the training phase:
\begin{align}
    \pi_\epsilon'(\boldsymbol{x}) =
        \begin{cases}
                \arg\max_{u} q_\pi(\boldsymbol{x}, u, \bm{\theta})   & \text{with probability}~1-\epsilon, \\
                \mathcal{U}_\mathcal{A}                   & \text{with probability}~\epsilon,
        \end{cases} 
        \label{eq:epsilon_greedy_policy}
\end{align}
with $\epsilon \in [0,1]$ being the randomization probability and $\mathcal{U}_\mathcal{A}$ denoting uniform random sampling from the finite action space $\mathcal{A}$. This exploration approach is prominently known as $\epsilon$-greedy policy \cite{SuttonBarto2015}.

%% file: chapters/Implementation.tex
\section{Implementation} \label{sec_Implementation}
\begin{figure}[htb]
    \centering
    \includegraphics[width=0.48\textwidth]{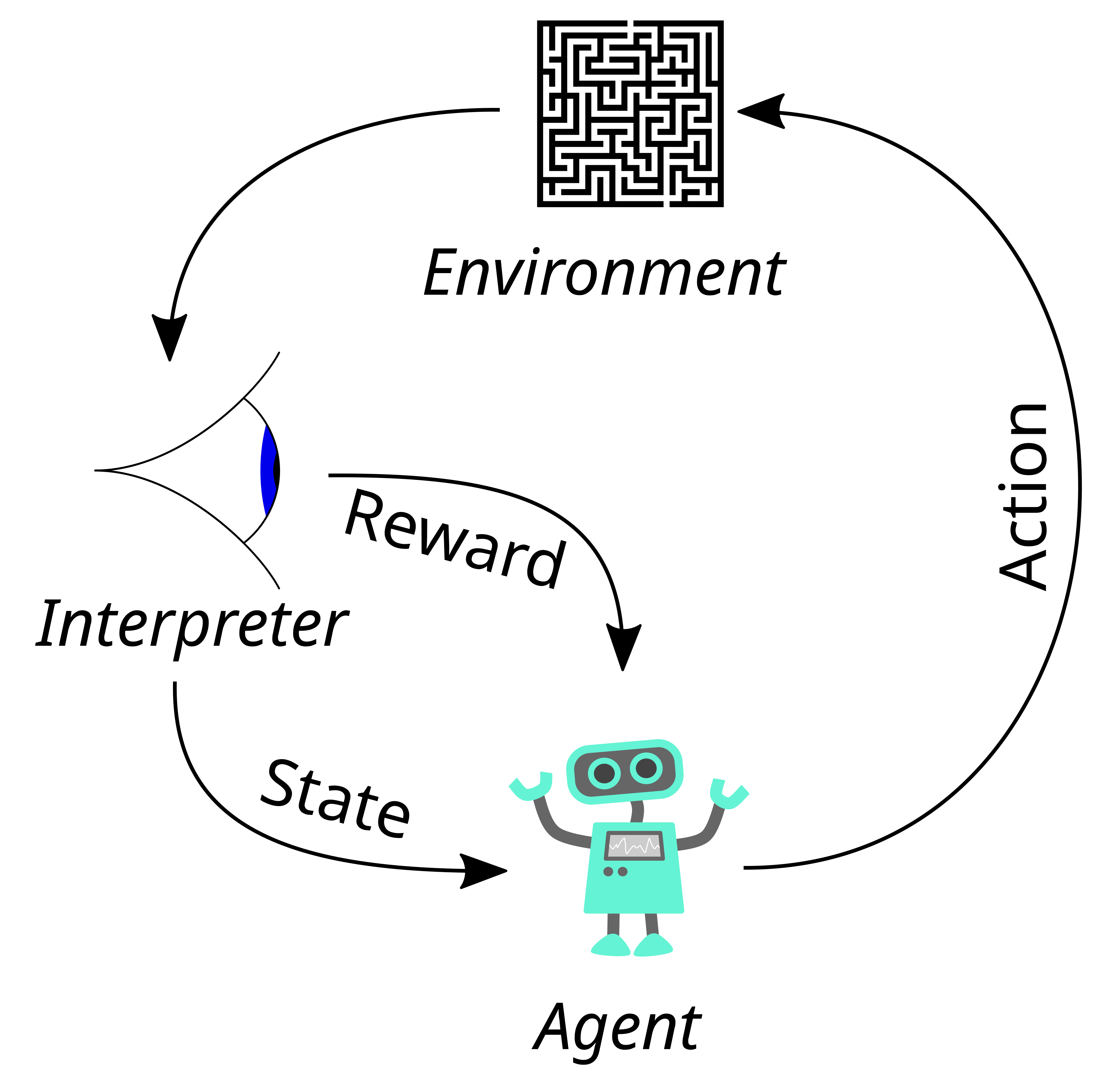}
    \caption{RL training loop \cite{megajuice_rl_diagram_2017}}
    \label{fig_RL_Training_Loop}
\end{figure}

The implementation of the simulation for the training environment and the adaptation and augmentation of the DQN are to be discussed in detail in the following. 
A schematic overview of the necessary elements that define an RL task can be seen in \figref{fig_RL_Training_Loop}.
Even though only a simplified unloading task is to be investigated in this work, the overall simulation scenario is designed such that it serves as a foundational framework for more complex tasks such as multi- or bulk-pick problems.

\subsection{Environment} 
\begin{figure*}[htb]
  \centering
  \includegraphics[width=\textwidth]{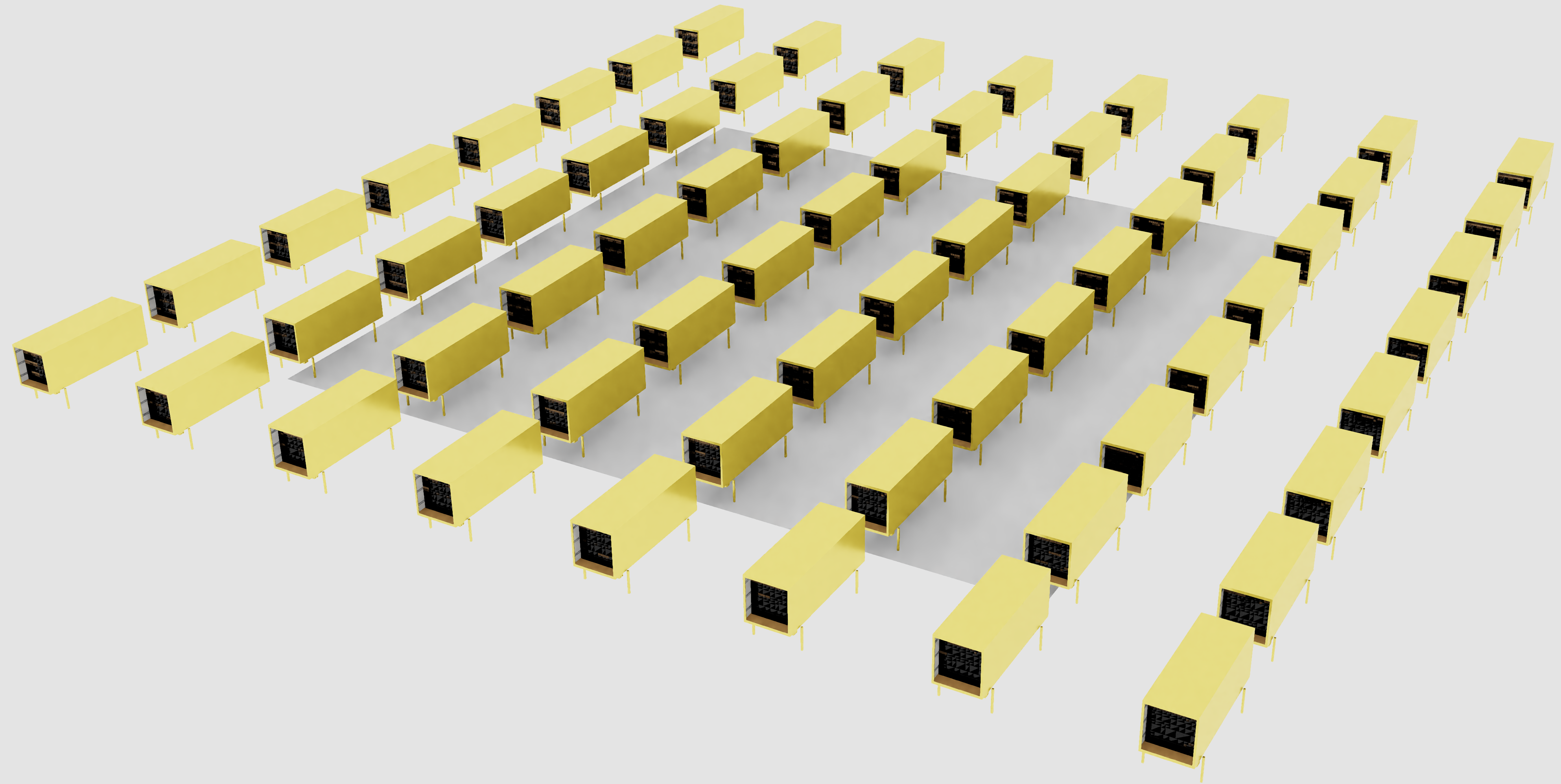}
  \caption{Simulation with 64 environments in parallel. Each environment contains one container and multiple items to be picked.
  }
  \label{fig_64envs}
\end{figure*}

The simulation environment is to mimic the physical behavior of the real world, whereas some simplifying assumptions have been defined for the proof of concept \secref{sec:introduction}.
Herein, the container itself is assumed a static, i.e., an immovable and rigid object. 
The packages are likewise assumed as rigid.
To accelerate the experience collection that takes place while interacting with this system, several environments are simulated in parallel, which is depicted in \figref{fig_64envs}.

\begin{figure*}[htb]
  \centering
  \begin{tikzpicture}[
    x=1cm, y=1cm,
    node distance=1.0cm and 1.0cm,
    startstop/.style = {ellipse, draw, minimum width=3cm, minimum height=1cm, align=center, fill=gray!20},
    process/.style = {rectangle, draw, align=center, fill=blue!15, text width=4cm,minimum width=4cm, minimum height=1cm},
    decision/.style = {diamond, draw, align=center, aspect=2 ,minimum width=4cm, minimum height=1.5cm, fill=gray!20},
    arrow/.style = {thick, ->, >=Stealth}
  ]
\node[draw, minimum width=1.8cm, minimum height=1.2cm, align=center, anchor=south west](1) at (0,1.8){1};
\node[draw, minimum width=3cm, minimum height=1.8cm, align=center, anchor=south west](2) at (0,0){0};
\node[draw, minimum width=1.8cm, minimum height=1.2cm, align=center, anchor=south west](3) at (3.6,1.8){1};
\node[draw, minimum width=1.2cm, minimum height=1.2cm, align=center, anchor=south west](4) at (5.4,1.8){2};
\node[draw, minimum width=3cm, minimum height=1.8cm, align=center, anchor=south west](5) at (4.8,0){0};
\node[draw, minimum width=1.8cm, minimum height=1.2cm, align=center, anchor=south west](6) at (8.4,1.8){1};
\node[draw, minimum width=1.2cm, minimum height=1.2cm, align=center, anchor=south west](7) at (10.2,1.8){2};
\node[draw, minimum width=1.8cm, minimum height=1.8cm, align=center, anchor=south west](8) at (9.6,0){3};
\node[draw, minimum width=1.8cm, minimum height=0.6cm, align=center, anchor=south west](9) at (0,6){8};
\node[draw, minimum width=0.9cm, minimum height=0.9cm, align=center, anchor=south west](10) at (0,5.1){6};
\node[draw, minimum width=0.9cm, minimum height=0.6cm, align=center, anchor=south west](11) at (0.9,5.4){7};
\node[draw, minimum width=0.9cm, minimum height=1.5cm, align=center, anchor=south west](12) at (0,3.6){5};
\node[draw, minimum width=2.1cm, minimum height=1.8cm, align=center, anchor=south west](13) at (0.9,3.6){4};
\node[draw, minimum width=1.8cm, minimum height=0.6cm, align=center, anchor=south west](14) at (3.6,6){8};
\node[draw, minimum width=1.2cm, minimum height=0.6cm, align=center, anchor=south west](15) at (5.4,6){9};
\node[draw, minimum width=1.2cm, minimum height=0.6cm, align=center, anchor=south west](16) at (3.6,5.4){9};
\node[draw, minimum width=0.9cm, minimum height=0.9cm, align=center, anchor=south west](17) at (4.8,5.1){6};
\node[draw, minimum width=0.9cm, minimum height=0.6cm, align=center, anchor=south west](18) at (5.7,5.4){7};
\node[draw, minimum width=0.9cm, minimum height=1.5cm, align=center, anchor=south west](19) at (4.8,3.6){5};
\node[draw, minimum width=2.1cm, minimum height=1.8cm, align=center, anchor=south west](20) at (5.7,3.6){4};
\node[draw, minimum width=1.2cm, minimum height=1.2cm, align=center, anchor=south west](21) at (8.4,5.4){2};
\node[draw, minimum width=0.9cm, minimum height=0.9cm, align=center, anchor=south west](22) at (9.6,5.7){6};
\node[draw, minimum width=0.9cm, minimum height=0.9cm, align=center, anchor=south west](23) at (10.5,5.7){6};
\node[draw, minimum width=1.8cm, minimum height=0.9cm, align=center, anchor=south west](24) at (9.6,4.8){10};
\node[draw, minimum width=1.8cm, minimum height=1.2cm, align=center, anchor=south west](25) at (9.6,3.6){1};
\node[draw, minimum width=0.6cm, minimum height=0.6cm, align=center, anchor=south west](26) at (0,9.6){12};
\node[draw, minimum width=0.6cm, minimum height=0.6cm, align=center, anchor=south west](27) at (0.6,9.6){12};
\node[draw, minimum width=0.6cm, minimum height=0.6cm, align=center, anchor=south west](28) at (1.2,9.6){12};
\node[draw, minimum width=0.9cm, minimum height=0.6cm, align=center, anchor=south west](29) at (0,9){7};
\node[draw, minimum width=0.9cm, minimum height=0.6cm, align=center, anchor=south west](30) at (0.9,9){7};
\node[draw, minimum width=1.5cm, minimum height=0.6cm, align=center, anchor=south west](31) at (0,8.4){11};
\node[draw, minimum width=1.5cm, minimum height=0.6cm, align=center, anchor=south west](32) at (1.5,8.4){11};
\node[draw, minimum width=1.8cm, minimum height=1.2cm, align=center, anchor=south west](33) at (0,7.2){1};
\node[draw, minimum width=1.2cm, minimum height=1.2cm, align=center, anchor=south west](34) at (1.8,7.2){2};
\node[draw, minimum width=0.6cm, minimum height=0.6cm, align=center, anchor=south west](35) at (3.6,9.6){12};
\node[draw, minimum width=0.6cm, minimum height=0.6cm, align=center, anchor=south west](36) at (4.2,9.6){12};
\node[draw, minimum width=0.6cm, minimum height=0.6cm, align=center, anchor=south west](37) at (4.8,9.6){12};
\node[draw, minimum width=0.6cm, minimum height=0.6cm, align=center, anchor=south west](38) at (5.4,9.6){12};
\node[draw, minimum width=0.6cm, minimum height=0.6cm, align=center, anchor=south west](39) at (6,9.6){12};
\node[draw, minimum width=1.2cm, minimum height=0.6cm, align=center, anchor=south west](40) at (3.6,9){9};
\node[draw, minimum width=0.9cm, minimum height=0.6cm, align=center, anchor=south west](41) at (4.8,9){7};
\node[draw, minimum width=0.9cm, minimum height=0.6cm, align=center, anchor=south west](42) at (5.7,9){7};
\node[draw, minimum width=1.5cm, minimum height=0.6cm, align=center, anchor=south west](43) at (4.8,8.4){11};
\node[draw, minimum width=1.5cm, minimum height=0.6cm, align=center, anchor=south west](44) at (6.3,8.4){11};
\node[draw, minimum width=1.8cm, minimum height=1.2cm, align=center, anchor=south west](45) at (4.8,7.2){1};
\node[draw, minimum width=1.2cm, minimum height=1.2cm, align=center, anchor=south west](46) at (6.6,7.2){2};
\node[draw, minimum width=0.9cm, minimum height=0.6cm, align=center, anchor=south west](47) at (8.4,9.6){7};
\node[draw, minimum width=0.6cm, minimum height=0.6cm, align=center, anchor=south west](48) at (9.3,9.6){12};
\node[draw, minimum width=0.6cm, minimum height=0.6cm, align=center, anchor=south west](49) at (9.9,9.6){12};
\node[draw, minimum width=0.9cm, minimum height=0.6cm, align=center, anchor=south west](50) at (10.5,9.6){7};
\node[draw, minimum width=0.6cm, minimum height=0.6cm, align=center, anchor=south west](51) at (8.4,9){12};
\node[draw, minimum width=1.2cm, minimum height=0.6cm, align=center, anchor=south west](52) at (9,9){9};
\node[draw, minimum width=1.2cm, minimum height=0.6cm, align=center, anchor=south west](53) at (10.2,9){9};
\node[draw, minimum width=0.9cm, minimum height=0.9cm, align=center, anchor=south west](54) at (9.6,8.1){6};
\node[draw, minimum width=0.9cm, minimum height=0.9cm, align=center, anchor=south west](55) at (10.5,8.1){6};
\node[draw, minimum width=1.8cm, minimum height=0.9cm, align=center, anchor=south west](56) at (9.6,7.2){10};
\node at (-0.9,1.5) {variant 2};
\node at (-0.9,5.1) {variant 1};
\node at (-0.9,8.7) {variant 0};
\node at (1.5,-0.6) {left};
\node at (6.3,-0.6) {mid};
\node at (10.2,-0.6) {right};
  \end{tikzpicture}
  \caption{Substack elements used to build walls of items. Each substack contains multiple package sizings (No.~0-12). Each item wall is composed of one left, two mid and one right substack.}
  \label{fig_basic_elements}
\end{figure*}
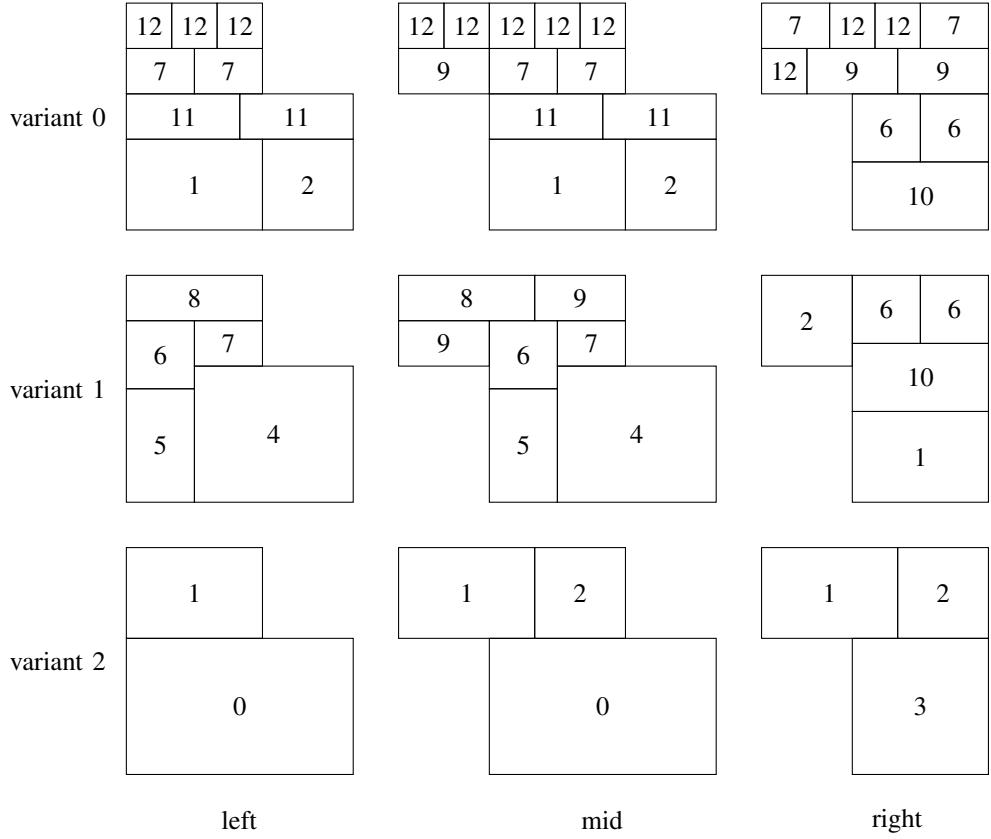

The pickable items are represented as simplified cardboard boxes of different sizes but with identical weight. 
In each container, a set of $800 - 1\,000$ items is placed in wall-like stacks, as it is common practice in the CEP industry. 
To prevent the agent from specializing for a specific stacking pattern, the items need to be positioned randomly to allow sufficient generalizability of the unloading policy.
Consequently, it is not desired to spawn the set of items identically within each of the environments, but individual stacking layouts are preferred.
These are to be composed of pre-defined package sizes, i.e., package dimensioning is not randomized, but may only take specific forms.
To create wall-like stacks of the specified sizings of packages, this work utilizes a set of predefined substack elements.
These substack elements ensure stackability of the composing packages (which not be trivial if package sizing and composition would be random) while keeping implementation effort low. 
Conceptually, an item wall is separated into three rows (top, middle (mid), bottom (bot)), with each row consisting of one left, two mid and one right substack element. All variants of substack elements are schematically depicted in \figref{fig_basic_elements}. 
 
The predefined substacks are then randomly combined to build a wall of items, with the restriction that a left substack can only be at the left side of the wall, a mid substack can only be in the middle of the item wall and a right substack can only be at the right side of an item wall. 
A total of 13 basic package sizings have been defined to compose the substacks and, eventually, the item walls. 
To additionally simplify the arrangement of consecutive item walls within the container, the packages only vary in terms of their width and height, but have identical spatial depth.

At start of a training episode, the substack variants are selected at random until each container is filled with seamless item walls. 
Caused by the random selection of substacks, the distribution of package sizings and the total number of items within the container varies from one environment to the next.
Programmatically, the initialization is handled by instantiating a large item selection globally at the synthesis of the 3D simulation, from which then the items can be drawn and moved to their intended container once the substacks are being determined. 
This way, the computational effort of registering new items within the simulation framework is avoided at the cost of instantiating more packages than needed.

\subsection{Observation Vector}\label{sub_sub_sec_Observations}
Next, the composition of the observation vector $\bm{x}$ is to be discussed.
As an interface between DQN and environment, it is of priority to equip the observation vector with all necessary information such that effective decision making is possible.
To keep the setup realistic, it is chosen to only consider the 128 items closest to the entrance of the container for the definition of $\bm{x}$, which is approximately the number of items that would be simultaneously visible for a human or camera. 
For simplification, the effects of obstructed view are herein not considered and only a distance measure is to be utilized.

Assuming that a human or robot operator would view the container from the top edge of the container entrance
\begin{align}
    \vec{p}_\text{view} &= 
    \begin{bmatrix}
        7.0~\text{m} & 1.25~\text{m} & 2.5~\text{m}
    \end{bmatrix}^\top.
\end{align}
The distance measure $d_i$ is computed according to the weighted Euclidean distance
\begin{align}
  d_i = \sqrt{(\vec{p}_\text{view} - \vec{p}_i)^\top \bm{W} (\vec{p}_\text{view} - \vec{p}_i)},
\end{align} 
the package index $i$,
the package position $\vec{p}_i$ and the weight matrix 
\begin{align}
  \bm{W} 
  = 
  \begin{bmatrix}
    1 & 0 & 0 \\
    0 & 0 & 0 \\
    0 & 0 & 4
  \end{bmatrix},
\end{align}
which is chosen such that the distance in y-direction (width of the container) is ignored and a difference in z-direction (height of the container) has importance over the distance in x-direction (depth of the container).  
This weighting leads to top-stack items being preferred over front-wall items for consideration in $\bm{x}$, which is pragmatic in a single-pick scenario.
This preprocessing targets a prioritized unloading of item wall closest to the container entrance, which is also realistic with respect of the limited arm length of a human or robot operator that could not feasibly reach packages that are positioned far within the container while the front wall is still blocking the path.
After sorting all items based on their distance to the viewer, only the 128 items with the smallest distance will be perceivable by the agent. 

Since it is assumed that the items are neatly stacked, their spatial rotation is uniform and does, hence, not carry sensible information. 
To simplify the learning task, the observations made available to the agent are therefore constructed from the item position only 
\begin{align}
    \vec{p}_{k, i} =
    \begin{bmatrix}
        p_{k, i, \text{x}} & p_{k, i, \text{y}} & p_{k, i, \text{z}}
    \end{bmatrix}^\top,
\end{align}
with $k$ denoting the time step at which the container is being observed.
Each item's position get scaled between $[-1, 1]$ with respect to the container's dimensions by
\begin{align}
     \tilde{\vec{p}}_{k, i}    = 
    \begin{bmatrix}
        \tilde{p}_{k, i, \text{x}} \\
        \tilde{p}_{k, i, \text{y}} \\
        \tilde{p}_{k, i, \text{z}}
    \end{bmatrix} =
     \begin{bmatrix}
        \tfrac{2}{7.0~\text{m}} & 0 & 0 \\
        0 & \tfrac{2}{2.5~\text{m}} & 0 \\
        0 & 0 & \tfrac{2}{2.5~\text{m}}
    \end{bmatrix}
    \vec{p}_{k,i}
    -
    \begin{bmatrix}
        1 \\ 1 \\ 1
    \end{bmatrix}.
\end{align}
Correspondingly, the observation space is 
\begin{align}
    \mathcal{O} \in \mathbb{R}^{128 \times 3},     
\end{align}
where each observation $\bm{x}_k$ is build from the position of each item at the respective time step $k$:
\begin{align}
    \boldsymbol{x}_k 
    = 
    \begin{bmatrix*}[l]
        \tilde{\vec{p}}^\top_{k, 0} \\
        \tilde{\vec{p}}^\top_{k, 1} \\
        \vdots\\
        \tilde{\vec{p}}^\top_{k, 127}
    \end{bmatrix*}.
    \label{eq:obs_def}
\end{align}

\subsection{Action}
After the agent has received the observation vector $\bm{x}$ as information about the state of the environment, it must act upon it by sensibly deciding on an action from the available action space $\mathcal{A}$, which is to be defined next.
Since this work simplifies the task to just select the item to be picked, the action space is discrete. 
It is composed of item indices $i$ of the 128 perceivable items as introduced in \secref{sub_sub_sec_Observations}:
\begin{align}
    \mathcal{A} = \{0, 1, \dots, 127\}.
\end{align}
The agent is to output a single item index which then specifies the item to be picked.

What follows after that, although outside the scope of the agents decisions, must be viewed and defined as an inherent part of the environment, i.e., the procedure that is applied to the selected item must be specified.
According to the simplifications discussed in \secref{sec:introduction}, the action execution reduces to application of a vertical force to the selected item. 
This force is fixed to $\vec{F}_\text{pick} = 20\,\text{N} \cdot \vec{e}_\text{z}$, which is applied at the center of mass of the selected item in positive z-direction\footnote{The symbols $\vec{e}_\text{x}$, $\vec{e}_\text{y}$ and $\vec{e}_\text{z}$ denote the positive unit vectors with respect to the x,y and z axis, i.e., $\vec{e}_\text{x}=[1 \, 0 \, 0]^\top$, $\vec{e}_\text{y}=[0 \, 1 \, 0]^\top$ and $\vec{e}_\text{z}=[0 \, 0 \, 1]^\top$.} over a time of $t = 0.3\,\text{s}$ in an attempt to lift the intended package. If the applied force is sufficient to move the selected item over a distance $\geq{d}_\text{threshold}=0.2\,\text{m}$, the item is considered unloadable and is immediately removed from the container. The numerical configuration of force, time and distance threshold ensures that stacks of two or more packages result to be unloadable, which is integral for the targeted single-pick task. A corresponding derivation is presented in Sec.~\hyperref[app:threshold_selection]{App.}.

Finally, the duration of a full training episode is defined as the agent being able to suggest $500$ actions. 
With the given item stacking algorithm, one container is initially loaded with $800-1\,000$ items, i.e., the agent will never be able to completely unload a container during one episode.
This way, it is ensured that there are always $>128$ items left such that the size of the observation and action space stays constant.

\subsection{Reward design} \label{sub_sub_sec_reward_design}
The main challenge within the unloading task relies in selecting pickable items consecutively. 
As per the discussion in the previous paragraph, unloadability of an item in a single pick fashion can be evaluated by comparing the traveled distance during a lifting attempt against the threshold of $d_\text{threshold}=0.2\,\text{m}$.
The traveled distance of the intended item is herein measured by an (unweighted) Euclidean distance
\begin{align}
    d_{k, i} = ||\vec{p}_{k, i} - \vec{p}_{k-1, i,} ||_2,
\end{align}
with the position of the $i$-th item $\vec{p}_{k-1, i}$ before the lifting attempt, and the position of the item $\vec{p}_{{k, i}}$ after the lifting attempt. 

This leads to the following binary reward design
\begin{align}
  r_k = \begin{cases}
    +1, &\text{if } d_i > d_\text{threshold},
    \\
    -1, & \text{otherwise},
  \end{cases}
\end{align}
i.e., a decision is viewed as successful if the selected item was movable, in which case the unloading is directly performed, or it is unsuccessful and has no effect upon the containers content.

\subsection{Discount factor $\gamma$}
Another yet unspecified degree of freedom is the value for the discount factor $\gamma$. 
Due to the structure of the task, it is feasible to assume that there are always pickable items available, and that the order of unloading does not play a role.
Thus, there is no evident benefit in a multi-step look-ahead strategy in the given single-pick scenario as later rewards are not critically affected by the momentary action.
Accordingly, the discount factor is configured to $\gamma = 0$ in this contribution, i.e., the decision problem is viewed in a one-step fashion. 
$\gamma$ being zero also is relevant for the loss function, because it entirely nullifies the need for a look-ahead action value $\hat{q}(\bm{x}_{k+1}, u', \bm{\theta}^-)$. 
Importantly, the occurrence of permutations within the observation vector $\bm{x}$ may break the temporal relation of observations over time, which would counteract any foresightedness.
An intuitive occurrence of this effect takes place when a package is being successfully unloaded, which may potentially lead to item coordinates in $\bm{x}$ changing positions while the coordinates of the unloaded item are removed, and the coordinates of a previously unseen item become visible.
Herein, it must be doubted whether the connection between prior and posterior observation is beneficial to the training. 

\subsection{Feature Engineering}
\label{sub_sec_feature_engineering}
\begin{figure}
\centering
\begin{tikzpicture}
    \tikzset{every x tick label/.append style={/pgf/number format/.cd, 
            set decimal separator={.},
            fixed,
            fixed zerofill,
            precision=1}} 
    \tikzset{every y tick label/.append style={/pgf/number format/.cd, 
            /pgf/number format/fixed,
            set decimal separator={.},
            fixed,
            fixed zerofill,
            precision=1}}
    \begin{axis}
        [
        width=0.38*\textwidth,
        height=0.38*\textwidth,
        scale only axis,
        domain=0:1,
        ymin=-1.1,ymax=1.1,
        xmin=-1.1,xmax=1.1,
        xtick={-1.0, -0.5, 0.0, 0.5, 1.0},
        ytick={-1.0, -0.5, 0.0, 0.5, 1.0},
        ytick distance={0.5},
        xtick distance={0.5},
        xlabel={$\tilde{p}_{0, i, \text{x}}$},
        ylabel={$\tilde{p}_{0, i, \text{z}}$}, 
        legend cell align=left, 
        legend style={draw=none, at={(0.5,1.0)}, anchor=south},
        legend columns = -1,
        grid=both,
        unbounded coords=discard,
        ]
        \addplot[only marks, mark=*, color= blue] table [x=x, y=y, col sep=comma] {data/obs_hist_equal/all_items.csv};
        \addlegendentry{All items \phantom{000}}
        \addplot[only marks, mark=*, color= red] table [x=x, y=y, col sep=comma] {data/obs_hist_equal/Observation.csv};
        \addlegendentry{Observed items}
    \end{axis}
\end{tikzpicture}
\caption{Side-view cross-section diagram of the item positions within the container. The entrance of the container is located on the right.
}
\label{fig_clusterd_obs}
\end{figure}
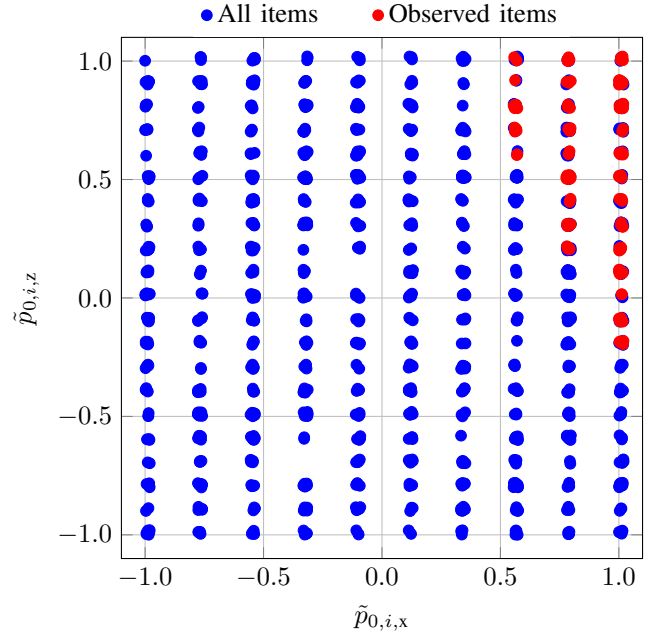

The fact that only the closest accessible 128 packages are perceived, as assumed within this work, has an effect on the coverage of the observation space.
As can be seen in \figref{fig_clusterd_obs}, the observed item positions are not uniformly distributed, whereas the blue marks represent all items within one container and the red ones represent the items considered visible. 
The grid-like structure herein originates from the well-ordered stacking. 
Based on the structure of the substacks and the pre-defined positions of packages with specific sizings, there is a limited number of possible positions that items can take within the container. 
In the given scenario, the observation space must hence be considered sparsely populated, for which a corresponding feature engineering (FE) can be beneficial to increase the resolution of observations.
Perturbation of the positions, which is visible as non-uniform blurring of the markers in \figref{fig_clusterd_obs}, originates from numerical inaccuracy of the simulation framework and has no systematic background in the given setup.
\\

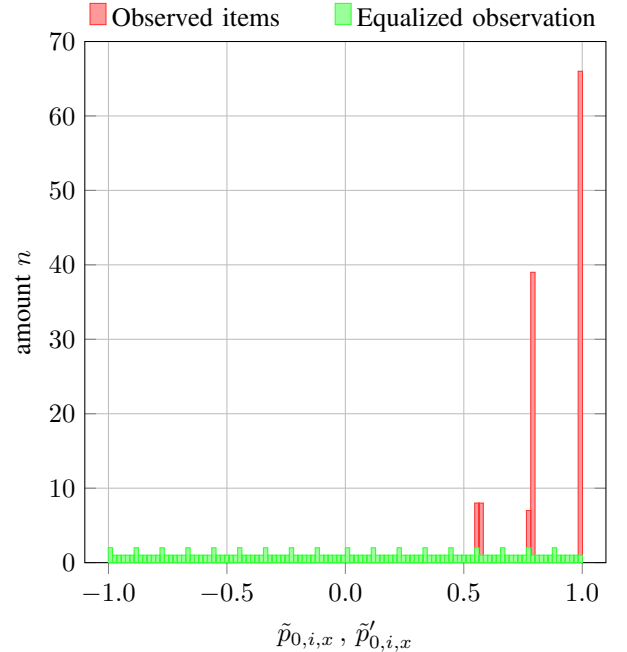
\begin{figure}[htb]
\centering
\begin{tikzpicture}
    \tikzset{every x tick label/.append style={/pgf/number format/.cd,
            set decimal separator={.},
            fixed,
            fixed zerofill,
            precision=1}}
    \tikzset{every y tick label/.append style={/pgf/number format/.cd,
            /pgf/number format/fixed,
            set decimal separator={.},
            fixed,
            fixed zerofill,
            precision=0}}
    \begin{axis}
        [
        width=0.38\textwidth,
        height=0.38*\textwidth,
        scale only axis,
        domain=-2:2,
        ybar, hist={data min=-1.001, data max=1.001},
        ymin=0, ymax=70,
        xmin=-1.1, xmax=1.1,
        ytick distance={10},
        xtick={-1.0,-0.5, 0, 0.5, 1.0},
        xlabel={$\tilde{p}_{0, i, x}\, ,\, \tilde{p}'_{0, i, x}$},
        ylabel={amount $n$},
        legend cell align=left, 
        legend style={draw=none, 
        at={(0.5,1.0)}, 
        anchor=south}, 
        legend columns = -1,
        grid=both,
        unbounded coords=discard,
        ]
        \addplot+[hist={bins=110, data min=-1.001, data max=1.001,}, fill=red!40, draw=red!80,] table [y=x, col sep=comma] {data/obs_hist_equal/Observation.csv};
        \addlegendentry{Observed items \phantom{000}}
        \addplot+[hist={bins=110, data min=-1.001, data max=1.001,}, fill=green!40, draw=green!80,] table [y=x, col sep=comma] {data/obs_hist_equal/Equalized_Observation.csv};
        \addlegendentry{Equalized observation}
    \end{axis}
\end{tikzpicture}
\caption{Histogram of the observed items' x-position $p_{0, i, \text{x}}$, at the beginning of a training episode, before histogram equalization in red and after histogram equalization in green.
}
\label{fig_obs_hist}
\end{figure}

To counteract the sparsity, a technique from image processing is adapted. 
When having a low contrast image, it is harder to process \cite{Laughlin1981}. 
Low contrast or sparsely populated observation spaces have a characteristically narrow histogram, where only a small part of the histogram contains observations, i.e., distribution of samples is concentrated in one area. 
To visualize this in \figref{fig_obs_hist} the histogram of the x-positions $p_{0, i, \text{x}}$ of the observed items is shown in red. 

\begin{figure}
  \begin{tikzpicture}
    \tikzset{every x tick label/.append style={/pgf/number format/.cd, 
            set decimal separator={.},
            fixed,
            fixed zerofill,
            precision=1}}
    \tikzset{every y tick label/.append style={/pgf/number format/.cd, 
            /pgf/number format/fixed,
            set decimal separator={.},
            fixed,
            fixed zerofill,
            precision=1}}
    \begin{axis}
        [
        width=0.38*\textwidth,
        height=0.38*\textwidth,
        scale only axis,
        domain=0:1,
        ymin=-1.1,ymax=1.1,
        xmin=-1.1,xmax=1.1,
        ytick distance={0.5},
        xtick distance={0.5},
        xlabel={$\tilde{p}'_{0, i, \text{x}}$},
        ylabel={$\tilde{p}'_{0, i, \text{z}}$},
        legend style={draw=none, at={(0.5,1.0)}, anchor=south},
        grid=both,
        unbounded coords=discard,
        ]
        \addplot[only marks, mark=*, color= green] table [x=x, y=y, col sep=comma] {data/obs_hist_equal/Equalized_Observation.csv};
    \end{axis}
\end{tikzpicture}
  \caption{Side-view cross-section diagram of the item positions after histogram equalization}
  \label{fig_not_clusterd_obs_hist_equal}
\end{figure}
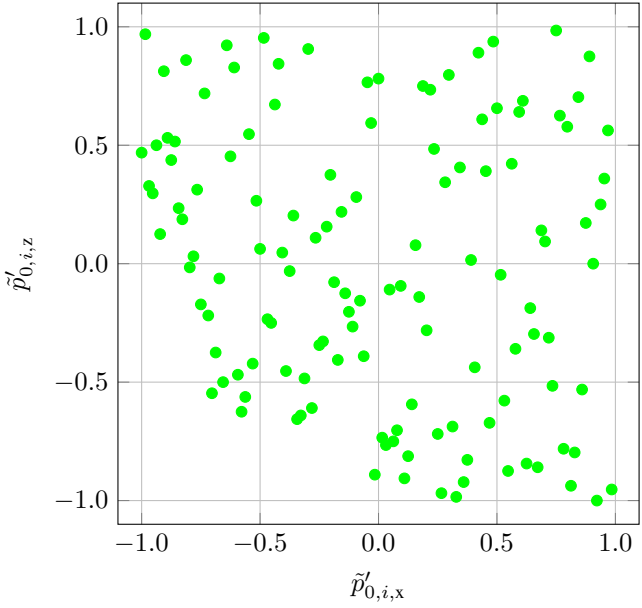

A compensation approach lies in the utilization of histogram equalization, which targets spreading out of a non-uniform sample distribution into a more uniform shape \cite{Laughlin1981}. 
Note that the original x-positions $p_{k, i, \text{x}}$ within \figref{fig_obs_hist} are not uniformly distributed whereas the new equalized x-positions $x_\text{new}$ are uniformly distributed. 
The histogram equalization procedure is performed on all three axes. 
Herein, the items remain in the same order, which is important for the item picking task such that, e.g., items on top of each other still have a higher z-position than the items below. 
An exemplary result of this form of FE can be seen in \figref{fig_not_clusterd_obs_hist_equal}, wherein a visibly more even coverage of the xz-plane has been achieved. 
A case-specific pseudocode can be found in Alg. \ref{code_hist_equal}, wherein the application to the given positional features is considered. 
An numerical example of histogram equalization can be seen in \tabref{tab:hist_eq}.

\begin{algorithm}
  \caption{Observation equalization}
  \label{code_hist_equal}
  \begin{algorithmic}[1]
  \Require positional feature of all items in one dimension $\tilde{\boldsymbol{p}}_n$
    \Function{Histogram\_Equalization}{$\tilde{\boldsymbol{p}}_n$}
    \State sorted\_pos\_values, sorted\_pos\_index $\gets$ sort($\tilde{\boldsymbol{p}}_n$)
    \State $\tilde{p}'_n$ $\gets  \frac{2 \cdot \text{sorted\_pos\_index} - 127}{127}$
    \State \Return ($\tilde{p}'_n$)
    \EndFunction
  \end{algorithmic}
\end{algorithm}

\begin{table}[htb]
\centering
\caption{Exemplary histogram equalization procedure on observed positions $p_{k,i,\text{x}}$}
\fontsize{7.7pt}{8.1pt}\selectfont
\begin{tabular}{l rrrrrrr}
\hline
Original $p_{k, i, \text{x}}$  & 0.10 & 0.12 & 0.11 & 0.20 & 0.19 & 0.18 & 0.30 
\\
sorted\_pos\_index      & 1    & 3    & 2    & 6    & 5    & 4    & 7    
\\
New $p'_{k, i, \text{x}}$  & -1.00&-0.33 & -0.66& 0.66 & 0.33 & 0.00 & 1.00 
\\
\hline
\end{tabular}
\label{tab:hist_eq}
\end{table}

Perspectively, it must be considered that histogram equalization distorts spatial relations between items, which could become problematic if additional features such as item size plays a role for the unloading task.

\subsection{Masked DQN with Permutation Equivalence} \label{sub_sub_sec_masked_DQN}
For the item picking task, the DQN agent serves as a score estimator with respect to each individual item. 
The standard DQN as described in \secref{sub_sec_standard_DQN} serves as a strong foundation, however, due to the structure of the task some further adaptations to the default setup can be made to improve the results. 

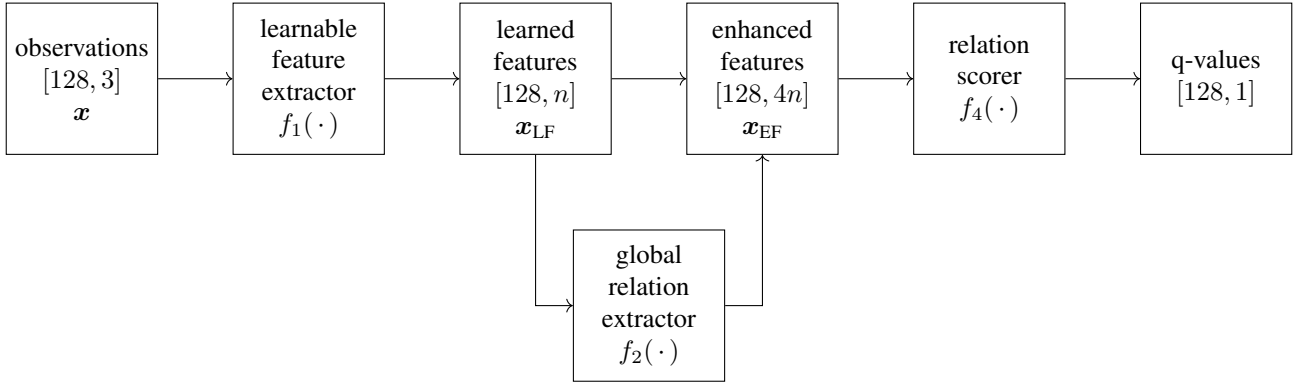
\begin{figure*}[htb]
\centering
\begin{tikzpicture}[
    x=1cm, y=1cm, yscale=-1,
    node distance=1.0cm and 1.0cm,
    startstop/.style = {ellipse, draw, minimum width=3cm, minimum height=1cm, align=center, fill=gray!20},
    process/.style = {rectangle, draw, align=center, fill=blue!15, text width=4cm,minimum width=4cm, minimum height=1cm},
    decision/.style = {diamond, draw, align=center, aspect=2 ,minimum width=4cm, minimum height=1.5cm, fill=gray!20},
    arrow/.style = {thick, ->, >=Stealth}
]
\node[draw,
      minimum width=2cm,
      minimum height=2cm,
      align=center,
      anchor=north west]
      (Obs) at (0,0)
      {observations \\ $[128,3]$ \\ $\boldsymbol{x}$};
\node[draw,
      minimum width=2cm,
      minimum height=2cm,
      align=center,
      anchor=north west]
      (LFE) at (3,0)
      {learnable\\
      feature \\ extractor\\ $f_1(\cdot )$};

\node[draw,
      minimum width=2cm,
      minimum height=2cm,
      align=center,
      anchor=north west]
      (Features) at (6,0)
      {learned \\ features \\ $[128, n]$ \\ $\boldsymbol{x}_\text{LF}$};
\node[draw,
      minimum width=2cm,
      minimum height=2cm,
      align=center,
      anchor=north west]
      (GRE) at (7.5,3)
      {global \\ relation \\ extractor \\ $f_2(\cdot )$};
\node[draw, minimum width=2cm, minimum height=2cm, align=center, anchor=north west](EF) at (9,0){enhanced \\ features \\ $[128, 4n]$ \\ $\boldsymbol{x}_\text{EF}$};
\node[draw,
      minimum width=2cm,
      minimum height=2cm,
      align=center,
      anchor=north west]
      (RS) at (12,0)
      {relation \\ scorer \\ $f_4(\cdot )$};
\node[draw,
      minimum width=2cm,
      minimum height=2cm,
      align=center,
      anchor=north west]
      (Q) at (15,0)
      {q-values \\ $[128, 1]$};   
\draw[->] (Obs) -- (LFE);
\draw[->] (LFE) -- (Features);
\draw[->] (Features.south) |- (GRE.west);
\draw[->] (Features) -- (EF);
\draw[->] (GRE.east) -| (EF.south);
\draw[->] (EF) -- (RS);
\draw[->] (RS) -- (Q);
\end{tikzpicture}
\caption{Overview of PEQ-network architecture.}
\label{PEQ_Highfly_Network_Architecture}
\end{figure*}

\subsubsection{Masking Actions to avoid Deadlocks}
If an item was selected despite not being pickable, it would be desired that the agent selects a different item in the next step. 
Due to the unsuccessful picking attempt, however, the position and orientation of items within the container does not change and, thus, the observation vector $\bm{x}$ remains the same. 
With the DQN-based selection procedure being deterministic and memoryless, another picking attempt would be started with respect to the same unavailable item.
The resulting loop would effectively halt the unloading procedure, i.e., it acts as a deadlock whose occurrence would be an exclusion criterion for the entire setup.
To address this risk, failed actions are memorized explicitly, allowing to ignore the unavailable item within the consecutive picking attempt.
This can be achieved by masking the DQN's output with respect to its action preference, i.e., the package corresponding to the second-largest action value is to be tried instead.
If necessary, this practice can be repeated until finally a pickable item has been reached, which would then reset the mask such that all items are approachable again. Note that this strategy is primarily required to manage non-ideal behavior during the application phase. In the training phase, the utilization of the randomized policy \eqref{eq:epsilon_greedy_policy} allows for different items to be attempted despite the observation remaining constant.  

\subsubsection{Architecture of the Deep Q-Network}
To design the DQN function approximator, the structure of the task and especially the observations need to be considered. 
In \secref{sub_sub_sec_Observations}, it was described that the 128 items closest to the entrance of the container are contained within $\bm{x}$. 
Although their positions are sorted by their distance to the entrance, it is only their relative position and not their order within the observation vector $\bm{x}$ (cf.~\eqref{eq:obs_def}) that should affect the action value.
Hence, permutation of two items within $\bm{x}$ should likewise exchange their computed action values $\hat{q}$, i.e., the DQN is required to be permutation equivalence. 
This results in the need of a specific network architecture that can guarantee permutation equivalence.
A regular fully connected multilayer perceptron is not generally permutation equivalent. 
Here, permutation equivalence only holds under specific conditions \cite{zaheer2017deepsets}.
A layer of a fully connected network is described by the function
\begin{align}
  \boldsymbol{y} = \sigma(\bm{\Theta}_i \boldsymbol{x})
\end{align}
with the input $\bm{x}$, the weight matrix $\bm{\Theta}_i$ of the $i$-th layer of a neural network, the activation function $\sigma(\cdot)$, which is a scalar function that is applied in element-wise fashion, and the output $\bm{y}$. 
If and only if $\bm{\Theta}_i$ satisfies
\begin{align}
  \bm{\Theta}_i = 
  \lambda_{i,1} \bm{I} + 
  \lambda_{i,2} (\boldsymbol{1}\boldsymbol{1}^\top), \label{eq_requirment_fully_connected_nn_permutation_equivalen}
\end{align}
with $\bm{I}$ being the identity matrix, $\bm{1} = [1\, 1\, \cdots \,1]^\top$ being a vector of ones and $\lambda_{i,1}$ and $\lambda_{i,2}$ being arbitrary scalars, the layer is permutation equivalent \cite{zaheer2017deepsets}. 
Utilizing random initialization in fully connected networks, the structure required according to \eqref{eq_requirment_fully_connected_nn_permutation_equivalen} is highly unlikely to occur and remain intact during gradient descent training.
Hence, fully connected networks are rather unsuitable for tasks that require permutation equivalence.

To ensure permutation equivalence, \cite{zaheer2017deepsets} suggests to utilize set operations on the observation vector, 
since for sets the order of elements is not relevant. 
If elements of a set are mutually exchanged, the set remains the same \cite{levin2018discrete}. 
Set operations are, e.g., $\max(\cdot)$, $\min(\cdot)$ or $\text{mean}(\cdot)$, which all carry the permutation invariance characteristic. 
To build a permutation equivalent network, a combination of 1D-convolutions and permutation invariant operations is suggested in the following. 
This architecture is schematically depicted in \figref{PEQ_Highfly_Network_Architecture}. 
Its main building blocks are the learnable feature extractor (LFE), the global relation extractor (GRE) and the relation scorer (RS). 
The inner structure of each element is described in the following. 

The LFE consists of 1D-convolutions, with a kernel size of $1 \times 3$, due to 3 features per item, a stride of $1\times1$ and no padding. 
Thus, the coordinates of each item get processed independently of other items and independent of its index within the observation matrix, rendering the operation permutation equivalent. 
The function $f_1(\bm{x}, \bm{\theta}_1)$ of the LFE is chosen as:
\begin{align}
  \bm{f}_{1,c} (\bm{x}_{k, i}, \bm{\theta}_{1, c}) 
  = 
  \sigma
  \left(\xi_{1, c} + 
  \sum_{j\in \{x, y, z \}}\bm{\Theta}_{1, c, j} 
  \bm{x}_{k, i, j}
  \right), 
  \label{eq_1d_conv}
\end{align} 
with the parameters $\bm{\theta}_{1, c} = \{\bm{\Theta}_{1,c}, \xi_{1, c}\}$, with $c \in \{0, 1, ..., n-1$, which compose $n$ kernels of shape $1\times3$, to yield $n$ learned features.
The activation is specified as ReLU function $\sigma(X) = \text{ReLU}(x) = \max(0, x)$. Therefore the LFE is described by
\begin{align}
    f_1(\boldsymbol{x}_{k,i}, \boldsymbol{\theta}_1)
    =
    \begin{bmatrix}
        f_{1,0}(\boldsymbol{x}_{k,i}, \boldsymbol{\theta}_{1,0}) \\
        f_{1,1}(\boldsymbol{x}_{k,i}, \boldsymbol{\theta}_{1,1})\\
        \vdots \\
        f_{1,n-1}(\boldsymbol{x}_{k,i}, \boldsymbol{\theta}_{1,n-1})
    \end{bmatrix}^\top.
\end{align}
The task of the LFE is, as the name already suggests, to learn which features are worth to consider. 
Note that the operation is fully permutation equivalent because the convolution is performed in a row-wise / item-wise fashion, determined by the selected values for the kernel size and stride and the utilization of (\ref{eq_1d_conv}). 
Therefore, the indexing order of items does not affect the calculated features.

After extracting meaningful features for each individual item, they need to be viewed in relation to other items, which is the task of the GRE. 
By using permutation invariant operations \cite{zaheer2017deepsets} such as min, max and mean over all items, the global context of each feature becomes available:
\begin{align}
    \bm{f}_2(\bm{x}_{\text{LF}}) &= 
    \begin{bmatrix}
        \min(\bm{x}_{\text{LF}}) &
        \max(\bm{x}_{\text{LF}}) &
        \text{mean}(\bm{x}_{\text{LF}})
    \end{bmatrix}^\top,
\end{align}
After the features of each item $\bm{x}_{\text{LF}}$ and the global context $\boldsymbol{x}_{\text{GC}}$ are computed, they are concatenated such that the item-specific features and their global maxima, minima and averages are available:
\begin{align}
    f_3(\bm{x}_{\text{LF}},\bm{x}_{\text{GC}}) 
    &= 
    \begin{bmatrix}
        \bm{x}_{\text{LF}} & \bm{x}_{\text{GC}}
    \end{bmatrix}^\top 
    = 
    \bm{x}_{\text{EF}}.
\end{align}

The enhanced features are subsequently processed by the "relation scorer" (RS), which computes the final q-values. 
Similar to the LFE, this is performed by pointwise convolutions applied independently to each item.
The RS is defined as
\begin{align}
     f_{4}(\boldsymbol{x}_{k,i}^{\text{EF}}, \boldsymbol{\theta}_2)
     =
     \tanh\!\left(
        \xi_{2}
        +
        \sum_{\ell=1}^{4n}
        \Theta_{2,\ell}\,
        x^{\text{EF}}_{k,i,\ell}
     \right),
\end{align}
with parameters $\boldsymbol{\theta}_2 = \{\boldsymbol{\Theta}_2, \xi_2\}$.
The kernel size is $1\times 4n$, a unit stride is used, and no padding is applied.
The activation function for the final layer is chosen such that $\sigma(x) = \tanh(x)$, which fits the expected range $[-1, 1]$ of the action values. 

Since the discount factor $\gamma = 0$, the target q-values are just the one step reward it can only be 1 or -1. 

\subsubsection{Further adaptations}
Instead of the loss function \eqref{eq_Standard_DQN_loss_fcn_MSE} according to \cite{Mnih2015}, this work utilizes the smooth-L1-loss \cite{SB3DQNDocs2026} for the DQN training: 
\begin{align}
\ell(a, b)&=
\begin{cases}
  \frac{(a - b)^2}{2\beta}, 
  & 
  \text{if }|a-b| < \beta
  \\
  |a - b| - \frac{\beta}{2}, & \text{otherwise} \label{eq_smooth_L1_loss}
\end{cases}, 
\end{align}
with $a$ being the momentary calculated action value estimate $\hat{q}(\boldsymbol{x}_k, u, \boldsymbol{\theta})$ and $b$ being the target $r + \gamma \max_u (\hat{q}(\boldsymbol{x}_{k+1}, u, \boldsymbol{\theta}^-))$. 
The threshold value is selected to $\beta = 1$. 
This loss function is less sensitive to outliers and is thus assumed to enable a more stable training process \cite{PyTorchSmoothL1Loss2026}. 

%% file: chapters/Results.tex
\section{Results}\label{sec_Results}
The training and application results are to be presented within this section. 
Before the training environment as described in \secref{sec_Implementation}, a less comprehensive scenario is analyzed for the optimization of the DQN's hyperparameters. 
After suitable hyperparameters have been found, they are used within the actual training environment to train the RL agent. 
The results of this training is described in \secref{sub_sec_training_results_masked_dqn_peq}.

\subsection{Hyperparameter Optimization}
\label{sub_sec_finding_hyperparameters}
Before the time-intensive main simulations are to be launched, it is of interest to find a suitable configuration of the DQN's hyperparameters. 
It should be noted that the performed hyperparameter optimization is deliberately limited in scope. Rather than aiming for a comprehensive or exhaustive search, the objective is to empirically identify a robust and reasonable hyperparameter configuration that allows stable training within the computational constraints of this work. The investigated parameter ranges are therefore restricted, and interactions between hyperparameters are not systematically explored. Consequently, the obtained configuration should be understood as practically suitable, but not necessarily optimal in a global sense.
Therefore, a hyperparameter-tuning environment is set up to generate training data quickly. 
It mimics the described training task (cf.~\secref{sec_Implementation}) without utilizing physically motivated data to saves computations.
According to the single-pick scenario, one simple but optimal policy would be to always pick the item on top of the item wall, i.e.,
to select the package with the highest positioning in z direction. 
To easily prototype on this premise, a random number generator is utilized to produce a randomized set of item coordinates as observation vector $\bm{x}$.
During training, the agent is then rewarded with $r=+1$ if it selects the item with the highest z coordinate, and with $r=-1$ otherwise.
Herein, it is ensured that the randomized coordinates are not arbitrary but match the grid-like structure that was already documented in \figref{fig_clusterd_obs}. 
The FE as of \secref{sub_sec_feature_engineering} and the further DQN extensions as of \secref{sub_sub_sec_masked_DQN} are not adapted and are utilized as described earlier. 

For better comparability to the later training in the simulation environment, instead of analyzing the reward directly, the mean reward is used to calculate the mean success rate (MSR), which is calculated by
\begin{align}
        \text{MSR} = \frac{r_\text{mean} + 1}{2}, \label{eq_mean_success_rate}
\end{align}
with the mean reward per pick $r_\text{mean}$.

A training is considered successful if an mean reward per pick close to $r_\text{mean}=+1$ is reached, which means that the agent always selects a pickable item. 
With this setup, different hyperparameter sets are to be tested. 
The following hyperparameters are tested in 10 runs each over $K = 3000$ training steps:
\begin{itemize}
  \item learning rate $\mu$ = $\{10^{-1}, 10^{-2}, 10^{-3}\}$,
  \item batch size $ b = \{64, 256, 1024, 2048\}$.
\end{itemize}
In \figref{fig_res_hyperparameter-tuning_hitrate_var_lr}, the mean success rate $\text{MSR}$ is shown with a constant batch size $b = 1024$ and different learning rates $\mu$. 
For a better visibility, only one representative curve per setup is shown, even though all curves show a similar behavior.
This figure shows that higher learning rates lead to unstable and in most cases unsuccessful training. Thus a learning rate $\mu = 10^{-3}$ is chosen for the later training.

\pgfplotsset{scaled y ticks=false}
\begin{figure}
\centering
\begin{tikzpicture}
    \tikzset{every x tick label/.append style={/pgf/number format/.cd,
            fixed,
            fixed zerofill,
            precision=0}}
    \tikzset{every y tick label/.append style={/pgf/number format/.cd,
            /pgf/number format/fixed,
            set decimal separator={.},
            fixed,
            fixed zerofill,
            precision=1}}
    \begin{axis}
        [
        width=0.4*\textwidth,
        height=0.4*\textwidth,
        scale only axis,
        domain=0:2500, 
        ymin=0,ymax=1,
        xmin=0,xmax=4000,
        ytick distance={0.2},
        xtick={0,1e3,2e3,3e3,4e3},
        xticklabels={$0$, $1\cdot10^{3}$, $2\cdot10^{3}$, $3\cdot10^{3}$, $4\cdot10^{4}$},
        ylabel={$\text{MSR}$},
        xlabel={Step $k$},
        legend style={draw=none, at={(0.5,1.0)}, anchor=south}, legend columns=-1,
        grid=both,
        unbounded coords=discard,
        ]
    \addplot[mark=only marks, color= blue] table [col sep=semicolon, y index=0, x expr=\coordindex] {data/dummy_sim_with_hist_equal/hitrate_curve_bz_1024_lr_0.1_agent_6.csv};
    \addlegendentry{$\mu = 10^{-1}$ \phantom{0}}
    \addplot[mark=only marks, color= red] table [col sep=semicolon, y index=0, x expr=\coordindex] {data/dummy_sim_with_hist_equal/hitrate_curve_bz_1024_lr_0.01_agent_7.csv};
    \addlegendentry{$\mu = 10^{-2}$ \phantom{0}}
    \addplot[mark=only marks, color= green] table [col sep=semicolon, y index=0, x expr=\coordindex] {data/dummy_sim_with_hist_equal/hitrate_curve_bz_1024_lr_0.001_agent_5.csv};
    \addlegendentry{$\mu = 10^{-3}$}
    \end{axis}
\end{tikzpicture}
\caption{Learning curve during hyperparameter tuning procedure with batch size $b=1024$ and different learning rates $\mu$.} 
\label{fig_res_hyperparameter-tuning_hitrate_var_lr}
\vspace{0.2cm}
\end{figure}
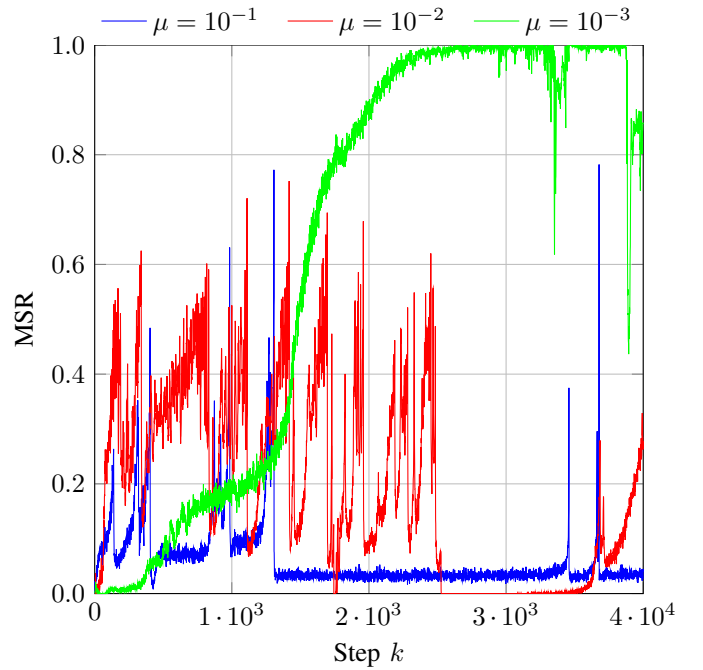

\begin{figure}
\centering
\begin{tikzpicture}
    \tikzset{every x tick label/.append style={/pgf/number format/.cd,
            fixed,
            fixed zerofill,
            precision=0}} 
    \tikzset{every y tick label/.append style={/pgf/number format/.cd, 
            /pgf/number format/fixed,
            set decimal separator={.},
            fixed,
            fixed zerofill,
            precision=1}}
    \begin{axis}
        [
        width=0.4*\textwidth,
        height=0.4*\textwidth,
        scale only axis,
        domain=0:2500,
        ymin=0,ymax=1,
        xmin=0,xmax=4000,
        ytick distance={0.2},
        xtick={0,1e3,2e3,3e3,4e3},
        xticklabels={$0$, $1\cdot10^{3}$, $2\cdot10^{3}$, $3\cdot10^{3}$, $4\cdot10^{4}$},
        ylabel={$\text{MSR}$},
        xlabel={Step $k$},
        legend style={draw=none, at={(0.5,1.01)}, anchor=south, legend cell align=left},
        legend columns=2,
        grid=both,
        unbounded coords=discard,
        ]
\addplot[mark=only marks, color= blue] table [col sep=semicolon, y index=0, x expr=\coordindex] {data/dummy_sim_with_hist_equal/hitrate_curve_bz_64_lr_0.001_agent_9.csv};
\addlegendentry{$b = 64$}
\addplot[mark=only marks, color= red] table [col sep=semicolon, y index=0, x expr=\coordindex] {data/dummy_sim_with_hist_equal/hitrate_curve_bz_256_lr_0.001_agent_9.csv};
\addlegendentry{$b = 256$}
\addplot[mark=only marks, color= green] table [col sep=semicolon, y index=0, x expr=\coordindex] {data/dummy_sim_with_hist_equal/hitrate_curve_bz_1024_lr_0.001_agent_5.csv};
\addlegendentry{$b = 1024$ \phantom{000}}
\addplot[mark=only marks, color= orange] table [col sep=semicolon, y index=0, x expr=\coordindex] {data/dummy_sim_with_hist_equal/hitrate_curve_bz_2048_lr_0.001_agent_6.csv};
\addlegendentry{$b = 2048$}
    \end{axis}
\end{tikzpicture}
\caption{Learning curve during hyperparameter tuning procedure with different batch sizes $b$ and constant learning rate $\mu=10^{-3}$.} 
\label{fig_res_hyperparameter-tuning_hitrate_var_bs}
\vspace{0.2cm}
\end{figure}
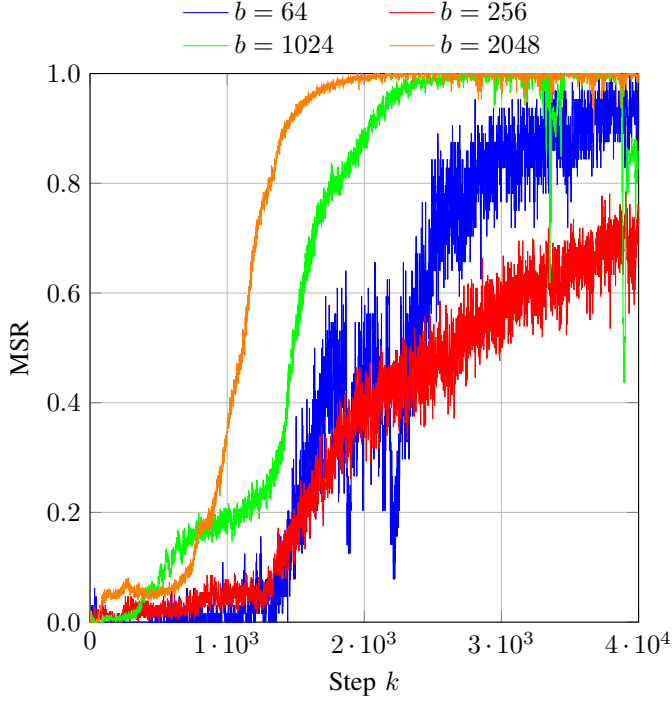

In \figref{fig_res_hyperparameter-tuning_hitrate_var_bs}, the above-mentioned hyperparameter setups are compared with a constant learning rate $\mu = 10^{-3}$ and different batch sizes $b$. 
For better visibility, only one representative curve per setup is shown. 
The observed training behavior indicates that a bigger batch size improves performance. 
Not only is the desired $\text{MSR}$ reached faster, the training is also less noisy and, thus, more stable. 
Therefore, a batch size $b=2048$ is chosen for the training in the simulated environment.

\begin{figure}
\centering
\begin{tikzpicture}
    \tikzset{every x tick label/.append style={/pgf/number format/.cd,
            fixed,
            fixed zerofill,
            precision=0}}
    \tikzset{every y tick label/.append style={/pgf/number format/.cd, 
            /pgf/number format/fixed,
            set decimal separator={.},
            fixed,
            fixed zerofill,
            precision=1}}
    \begin{axis}
        [
        width=0.4*\textwidth,
        height=0.4*\textwidth,
        scale only axis,
        domain=0:2500,
        ymin=0,ymax=0.6,
        xmin=0,xmax=4000,
        ytick distance={0.1},
        xtick={0,1e3,2e3,3e3,4e3},
        xticklabels={$0$, $1\cdot10^{3}$, $2\cdot10^{3}$, $3\cdot10^{3}$, $4\cdot10^{4}$},
        ylabel={$\ell$}, 
        xlabel={Step $k$},
        legend style={draw=none, at={(0.5,1.0)}, anchor=south},legend columns=-1, 
        grid=both,
        unbounded coords=discard,
        ]
        \addplot[mark=only marks, color= blue] table [col sep=semicolon, y index=0, x expr=\coordindex] {data/dummy_sim_with_hist_equal/loss_curve_bz_2048_lr_0.001_agent_6.csv};
        \addlegendentry{with FE \phantom{000}}
        \addplot[mark=only marks, color= red] table [col sep=semicolon, y index=0, x expr=\coordindex] {data/dummy_sim_without_hist_equal/loss_curve_bz_2048_lr_0.001_agent_8.csv};
        \addlegendentry{without FE}
    \end{axis}
\end{tikzpicture}
\caption{Training loss over time using the hyperparameter-tuning environment.} 
\label{fig_res_hyperparameter-tuning_loss_w_wo_fe}
\vspace{0.2cm}
\end{figure}
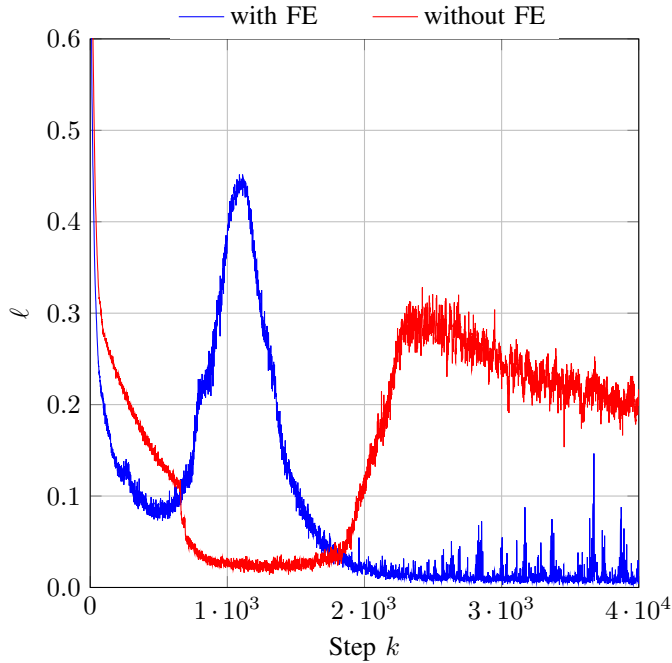

In \figref{fig_res_hyperparameter-tuning_loss_w_wo_fe}, the curve of the loss function $\ell$ described in (\ref{eq_smooth_L1_loss}) is shown for the chosen hyperparameter setup with and without the FE described in \secref{sub_sec_feature_engineering}. 

Note that, in both cases, the loss function first decreases before increasing again. 
Then, after reaching a peak, the loss function decreases again. 
Apparently, this is when the agent starts to learn how to select pickable items. 
In the case without FE, the loss function behaves similarly but the training takes significantly longer and is more unstable. 
This proves the benefit of proper FE in general, and of the proposed FE setup in particular. 

\subsection{Training Results masked DQN with PEQ agent}
\label{sub_sec_training_results_masked_dqn_peq}
With the result of the empirical hyperparameter optimization described in the previous section, the following configuration is utilized to setup the training with the physics-based simulation environment:
\begin{itemize}
    \item training steps $K = 20 \cdot 10^6$,
    \item learning rate $\mu = 10^{-3}$, 
    \item batch size $b = 2048$, 
    \item initial exploration rate $\epsilon_\text{init} = 1$, 
    \item final exploration rate $\epsilon_\text{final} = 0$
    \item exploration rate decrease interval $K_\epsilon = \frac{K}{2}$ (starting from $k=0$), 
    \item size of replay buffer: $2^{20}$.
\end{itemize}

\begin{figure}
\centering
\begin{tikzpicture}
    \tikzset{every x tick label/.append style={/pgf/number format/.cd,
            fixed,
            fixed zerofill,
            precision=1}} 
    \tikzset{every y tick label/.append style={/pgf/number format/.cd, 
            /pgf/number format/fixed,
            set decimal separator={.},
            fixed,
            fixed zerofill,
            precision=1}} 
    \begin{axis}
        [
        width=0.45*\textwidth, 
        height=0.45*\textwidth,
        domain=0:2500, 
        ymin=0,ymax=0.6, 
        xmin=0,xmax=20000000,
        ytick distance={0.1}, 
        ylabel={$\ell$},
        legend style={draw=none, at={(0.5,1.0)}, anchor=south}, 
        grid=both,
        unbounded coords=discard,
        scaled x ticks = false,
        xtick={0,5e6,10e6,15e6,20e6},
        xticklabels={$0$, $5\cdot10^{6}$, $10\cdot10^{6}$, $15\cdot10^{6}$, $20\cdot10^{6}$},
        xlabel={Step $k$},
        ]
        \addplot[mark=only marks, color= blue] table [x=Step, y=Value, col sep=comma] {data/loss_function_agent.csv};
    \end{axis}
\end{tikzpicture}
\caption{Exemplary training loss of the agent when training with the physics-based simulation environment.} 
\label{fig_res_loss_agent}
\vspace{0.2cm}
\end{figure}
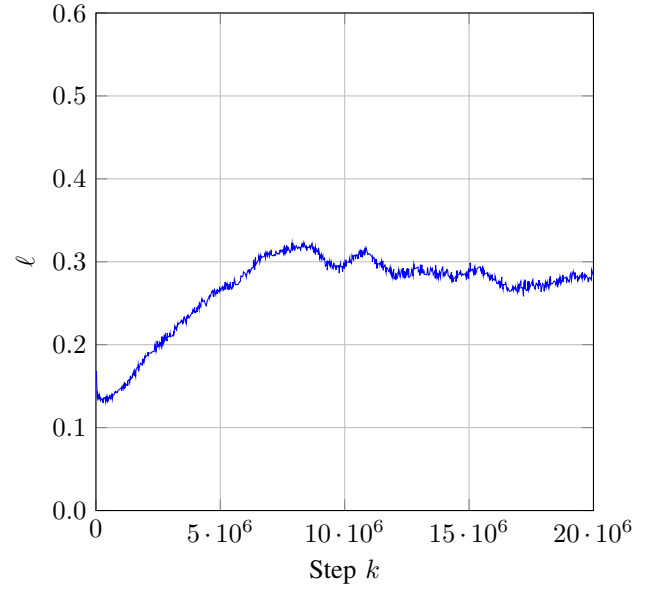

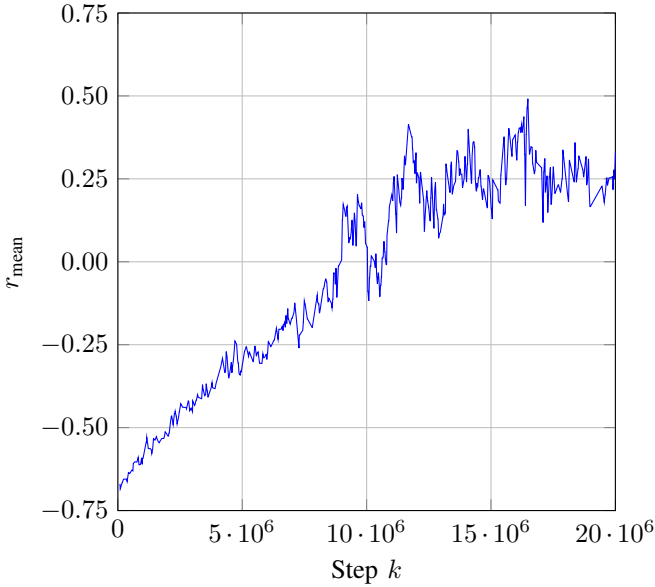
\begin{figure}
\centering
\begin{tikzpicture}
    \tikzset{every x tick label/.append style={/pgf/number format/.cd, 
            fixed,
            fixed zerofill,
            precision=1}} 
    \tikzset{every y tick label/.append style={/pgf/number format/.cd, 
            /pgf/number format/fixed,
            set decimal separator={.},
            fixed,
            fixed zerofill,
            precision=2}} 
    \begin{axis}
        [
        name=leftaxis,
        width=0.45*\textwidth,
        height=0.45*\textwidth,
        domain=0:20000000, 
        ymin=-0.75,ymax=0.75,
        xmin=0,xmax=20000000, 
        ytick distance={0.25},
        xtick={0,5e6,10e6,15e6,20e6},
        xticklabels={$0$, $5\cdot10^{6}$, $10\cdot10^{6}$, $15\cdot10^{6}$, $20\cdot10^{6}$},
        ylabel={$r_\text{mean}$},
        xlabel={Step $k$},
        legend style={draw=none, at={(0.5,1.0)}, anchor=south},
        grid=both,
        unbounded coords=discard,
        scaled x ticks = false,
        xlabel={Step $k$},
        ]
        \addplot[mark=only marks, color= blue,] table [x=Step, y expr={\thisrow{Value} / 500}, col sep=comma] {data/total_rew_function_agent.csv};
    \end{axis}
        anchor=south west,    
\end{tikzpicture}
\caption{Ensemble mean reward of 64 simultaneous environments over training time. 
} 
\label{fig_res_total_rew_agent}
\vspace{0.2cm}
\end{figure}

The training is conducted in an episodic manner, wherein the environment is regularly reset to a fully loaded container.
Each training episode gets terminated automatically after 500 steps to ensure that always more than 128 items are left in the container. The resulting loss curve is shown in \figref{fig_res_loss_agent} and the mean reward per episode is shown in \figref{fig_res_total_rew_agent}. 

Similarly as observed during hyperparameter investigation, the training loss decreases at first and then increases again. 
However, the decrease has a much lower slope, i.e., the negative trend is significantly less significant and more noisy than in the hyperparameter-tuning environment.  
In contrast to the hyperparameter-tuning environment, not only the item with the highest z-coordinate is pickable, but a selection of items is equally well accessible, leading to the optimal picking strategy being non-unique. 

There are two different possible explanations for the overall trend of the loss curve. 
One explanation is that, until the first minimum, the network just generally tunes its weights to be an appropriate value range. 
After that, the actual fitting to the training data starts. 
Another explanation could be, that the agent first learns the majority class, in case of the item picking task the non-pickable items. 
Therefore, an easy guess is to calculate an action value of $\hat{q}=-1$ for all items which is correct in about $80\,\%$ of all cases. 
This happens until the first minimum. 
After that the agent learns to distinguish between pickable an non-pickable items. 
At first this increases the loss function but also the reward, before finally the loss function decreases as well. 

In \figref{fig_res_total_rew_agent}, the mean reward per step $r_\text{mean}$ is shown. 
At the start of the training, the agent picks random items which lead to an mean reward $-0.7$, corresponding to a mean success rate of $\text{MSR} = 15\,\%$ according to \eqref{eq_mean_success_rate}. 
After that, a steady increase of the mean reward can be seen, although the increase of mean the reward becomes slower and more noisy after the loss function reaches its peak. 
Finally, it ends at around $0.2$, which corresponds to $\text{MSR} = 60~\%$. 
For comparison, the chance to select a pickable item at random is approximately $20~\%$, because due to the stacking algorithm about $20~\%$ of all items have no item on top and are therefore pickable. 
This is where the initial success rate comes from. 
An optimal policy achieves a $\text{MSR} = 100~\%$, so a successful pick every attempt. 
Therefore, the trained agent is significantly better than random, but still far from perfect.

%% file: chapters/Summary.tex
\section{Conclusion and Outlook} 
\label{sec_Conclusion_and_Outlook}
In this work, a fundamental framework was presented to examine whether the single-pick container unloading task can be solved utilizing RL. 
For that, a basic simulation environment was defined with the premise of being reusable in later studies targeting e.g., multi-pick strategies or unloading of orderless stacks.
A permutation equivalent FE and observation preprocessing provides the agent with sensible observations
while a masked DQN structure allows the avoidance of deadlocks that can be the result of suboptimal policies. 
The training process of the agent was presented and discussed, whereas a mean success rate of $60~\%$ was reported.
This first result is obviously far from perfect, but it is significantly better than a random policy. 
Hence, it is concluded that the item unloading task is principally learnable by means of RL methods, with potential improvements being discussed in the following. 

\subsection{Outlook}\label{sec_outlook}
In contrast to the initial expectations, the development of an effective solution still requires substantial expert knowledge, particularly in feature engineering, network design, and reward shaping. Rather than considering these aspects as finalized, they open up several avenues for further investigation. The following points outline possible directions for future work:

\begin{itemize}
\item \textbf{Scenario difficulty:}  
The current setup could be extended to more challenging scenarios, such as bulk-picking tasks or situations without well-structured item stacks, in order to assess robustness under less constrained conditions.
\item \textbf{Real-world interfaces:}  
The use of ground-truth positional information from the simulation could be replaced by real-world sensing systems. This includes the integration of cameras, perception pipelines, and localization algorithms, as well as a robotic arm.
\item \textbf{Robotic manipulation:}  
Adding a robotic arm would require not only identifying pickable items but also determining precise grasp points, forces, trajectories, and suitable grippers.
\item \textbf{Sim-to-real transfer:}  
With more realistic perception and actuation models, the feasibility of transferring policies from simulation to real-world systems can be investigated.
\item \textbf{Network architecture:}  
The current network design is still relatively limited. Exploring more expressive architectures may improve generalization and performance in complex scenarios.
\item \textbf{Hyperparameter optimization:}  
A more comprehensive and systematic hyperparameter optimization, conducted at a larger scale than in the present work, could further improve learning stability and performance.
\item \textbf{Reward function design:}  
Alternative reward formulations could be explored, for example assigning intermediate rewards to items positioned higher in the stack even if they are not yet pickable, to better guide the learning process.
\end{itemize}

%% file: chapters/Appendix_Derivation.tex
\section*{Appendix \\ Unloadability Classification}
\addcontentsline{toc}{section}{Appendix: Unloadability Classification}
\label{app:threshold_selection}
The following considerations are delivered to briefly motivate the configuration of $F_\text{agent}=20\,\text{N}$, $T_\text{lift}=0.3\,\text{s}$ and $d_\text{threshold}=0.2\,\text{m}$ for classifying whether an item can be unloaded or not. By Newton's second law, acceleration evaluates to
\begin{align}
  \vec{F} &= m \, \vec{a} &&\Leftrightarrow&  \vec{a} &= \frac{\vec{F}}{m}, 
\end{align}
where $m$ is the accelerated mass and $\vec{F}$ the sum of all forces upon the mass.
Note that both, the externally applied force $\vec{F}_\text{agent}$ and the gravitational force $\vec{F}_g$, act along the z axis. A single item's traveled distance $d$ then evaluates to
\begin{align}
  d&=|\vec{d}| = d_\text{z} = \frac{1}{2} a_\text{z} T_\text{lift}^2 
  = 
  \frac{1}{2} \frac{F_\text{z}}{m} T_\text{lift}^2 
  = 
  \frac{1}{2}\cdot \left(\frac{F_\text{agent}-F_g}{m}\right) \cdot T_\text{lift}^2, \nonumber\\
   &= \frac{1}{2} \left(\frac{F_\text{agent}}{m} - g \right) \cdot T_\text{lift}^2 
   \text{ if }\frac{F_\text{agent}}{m}>g,
   d=0 \text{ otherwise}
   , 
\end{align}
with the time $T_\text{lift}$ being the time duration for which lifting is attempted.
The lifting distance for lifting $p$ packages simultaneously results to
\begin{align}
     d_p &= \frac{1}{2} \left(\frac{F_\text{agent}}{p\cdot m} - g \right)\cdot T_\text{lift}^2
     \text{ if }\frac{F_\text{agent}}{p\cdot m} > g,
     d_p=0 \text{ otherwise}.
\end{align}
Plugging in the specifications $F_\text{agent}=20\,\text{N}$, $T_\text{lift}=0.3\,\text{s}$ and $m=1\,\text{kg}$ yields
\begin{align}
    d_{p=1} &= 0.46\,\text{m},
    &
    d_{p=2}&=9\,\text{mm},
\end{align}
wherein it is consistently assumed that all items are at rest before lifting is attempted, i.e., initial velocity is always zero.
With this setup, setting the distance threshold to \mbox{$d_\text{threshold} = 0.2\,\text{m}$} provides a clear and simple decision rule that allows to classify whether the selected item is unloadable. This primes the learned policy for single picks by disallowing removal of entire stacks of two and more items.